# Reinforcement Learning for Corporate Bond Trading: A Sell Side Perspective (Working Paper)

Samuel Atkins*  
U.S. Bank AI Research

Ali Fathi*  
U.S. Bank AI Research

Sammy Assefa*  
U.S. Bank AI Research

**Abstract**

A corporate bond trader in a typical sell side institution such as a bank provides liquidity to the market participants by buying/selling securities and maintaining an inventory. Upon receiving a request for a buy/sell price quote (RFQ), the trader provides a quote by adding a spread over a *prevalent market price*. For illiquid bonds, the market price is harder to observe, and traders often resort to available benchmark bond prices (such as MarketAxess, Bloomberg, etc.). In [BG23], the concept of *Fair Transfer Price* for an illiquid corporate bond was introduced which is derived from an infinite horizon stochastic optimal control problem (for maximizing the trader's expected P&L, regularized by the quadratic variation). In this paper, we consider the same optimization objective, however, we approach the estimation of an optimal bid-ask spread quoting strategy in a data driven manner and show that it can be learned using Reinforcement Learning. Furthermore, we perform extensive outcome analysis to examine the reasonableness of the trained agent's behavior.

## 1 Introduction

At the start of the year 2024, in the U.S. fixed income market, about 43% of investment-grade (IG) corporate bonds and 30% of high-yield (HY) corporate bonds were traded electronically [1]. Compared to 25% of IG and 13% of HY bonds which were traded electronically in early 2019, this shows a significant *"electronification"* trend in the post-COVID years. Compared to the equities, this trend has been somehow delayed in the fixed income market due to fundamental differences such as *fragmentation* of securities as well as differences in the types of market participants. For instance, according to SIFMA, there are around 66,000 U.S. corporate bonds that are actively traded, compared to about 4,500 stocks, as companies usually issue a variety of bonds with different profiles. Also, unlike the equity market, where retail investors have a prominent presence, the fixed income market is dominated by institutional investors for most asset classes.

---

*The views and opinions expressed in this article are solely those of the authors and do not reflect the views of or practices within U.S. Bank or any of its affiliates, including without limitation U.S. Bancorp Investments, Inc.

[1] According to Coalition Greenwich- an analytics company: https://www.greenwich.com/market-structure-technology/february-spotlight-record-start-2024-corporate-bonds

Market makers such as banks play an important role in the corporate bond market. Higher fragmentation of the asset class means typically lower trading activity at the individual bond (uniquely identified by CUSIP) level. From a liquidity perspective, bond buyers are rarely present when sellers want to sell and vice versa. Broker-dealers provide additional liquidity to market participants by buying bonds from sellers and holding them in inventory until they are matched with buyers. In short, the objective of a corporate bond trader is to provide liquidity to market participants by holding inventory while mitigating inventory risk and making a profit on the bid-ask spread.

**The RFQ Protocol: A Sell Side's View**   According to SIFMA, about 60% of corporate bond electronic trading is conducted via the request for quote (RFQ) protocol. At a high level, this is how a corporate bond trader in a bank receives electronic RFQs. The RFQs show up on the trader terminal queue with a ticking time to expiry of a few minutes. At this point, the trader considers the parameters of the RFQ, including the counterparty, the bond CUSIP, and available third-party benchmark prices (such as MarketAxess CP+, Bloomberg BVAL or other third parties). Then, she combines this information with her intuition about the market, and provides a response to the RFQ. It may happen that the counterparty cancels the RFQ or the response is not submitted by the trader within the time window, which results in the expiration of the RFQ. For each RFQ, the trader is optimizing the trade-off between responding within a short time period, providing a quote with the highest chance of winning the trade, and putting on a trade which aligns with the the desk's business strategy and risk appetite.

**Scope of this Paper**   This work is inspired by the notion of Fair Transfer Price (FTP) introduced by Bergault and Gueant in [BG23]. At a high level, the FTP framework is a pricing framework for a market maker who has access to patterns in the RFQ flow. The FTP framework has several desirable properties for pricing illiquid OTC securities (see section (3) for details). In this paper we focus on relaxing some of the *model-based* aspects of the framework and make them more *data-driven*. More precisely, we take the stochastic optimal control objective at the heart of the FTP and try to estimate the optimal control policy (response to bid-ask RFQs) via Reinforcement learning (RL). This is in contrast with [BG23], where a parametric dynamics model for the bond price is assumed and the optimal strategy is obtained by (numerically) solving the resulting PDE.

The RL approach does not need access to the dynamics model for the bond price and depends only on the step-wise reward (P&L as a result of answering RFQs at each step, penalized by the P&L variability) to learn the optimal strategy. Hence, a trading desk can train the RL agent on their RFQ historical data and use either the most recent TRACE prints (for the more liquid CUSIPs) or available third party benchmark composite prices (for the less liquid CUSIPs) as the reference bond price. In this paper, for the ease of external publication, we work with the model-based price dynamics and the calibrated parameters in [BG23].

While the FTP is calculated based on an *infinite horizon* stochastic optimal control problem (Equation (19)), our RL formulation is *finite horizon*. This is motivated by limits on the duration of holdings and positions as part of the Volcker Rule which limits principal trading activities by the banks for the purpose of market making [2].

The rest of the paper is organized as follows. In section 2, we review the mathematical background for the FTP framework. We also provide a brief overview of the RL algorithms we implemented. In section 3, we summarize the FTP framework in [BG23]. In Section 4, we describe the finite-horizon objective for an optimal strategy for answering bid-ask RFQs. Then, we train the RL algorithms and demonstrate training convergence. In section 6, we examine the trained RL agent via behavioral testing and scenario analysis.

## 2  Quantitative Preliminaries

In this section, we provide a review of concepts and tools used in the subsequent sections.

### 2.1  Stochastic Optimal Control and HJB Equations

Stochastic optimal control is at the heart of the FTP framework. We give a brief overview of the preliminaries here. We follow the exposition in [Pha09] closely. Throughout, we assume a filtration $(\Omega, \mathcal{F}, (\mathcal{F})_t$. Consider an SDE of the form

$$dX_s = b(X_s, s)ds + \sigma(X_s, s)dW_s, X_t = x_{\text{init}}. \tag{1}$$

on the time interval $s \in [t, T]$. The drift coefficient is given by $b : \mathbb{R}^d \times [t, T] \longrightarrow \mathbb{R}^d$ and $\sigma : \mathbb{R}^d \times [t, T] \longrightarrow \mathbb{R}^{d \times d}$ denotes the diffusion coefficient, $W_s$ denotes standard d-dimensional Brownian motion, and $x_{\text{init}} \in \mathbb{R}^d$ is the (deterministic) initial condition. We assume the "usual" regularity conditions for the drift and the diffusion coefficients.

A *controlled SDE* is given by,

$$dX_s^u = b(X_s, u_s, s)ds + \sigma(X_s, u_s, s)dW_s, X_t = x_{\text{init}}. \tag{2}$$

where we think of $u : \mathbb{R}^d \times [t, T] \longrightarrow \mathbb{R}^d$ as an external input controlling the dynamics. For theoretical tractability, it is assumed that $u \in \mathcal{U}$, the set of admissible controls which entails smoothness, boundedness and appropriate growth conditions in $x$.

Let $c : \mathbb{R}^d \times \mathbb{R}^d \longrightarrow \mathbb{R}^+$ and $C : \mathbb{R}^d \times \mathbb{R}^d \longrightarrow \mathbb{R}^+$ be the running cost and the terminal cost, respectively. The *cost-to-go* functions are defined as,

$$J^u(x, t) = \mathbb{E}\left[\int_t^T c(X_s^u, u_s)ds + C(X_T^u) \middle| X_t^u = x\right]. \tag{3}$$

The *value functions* $v : \mathbb{R}^d \times [0, T] \to \mathbb{R}^+$, are defined as,

$$v(x, t) = \inf_{u \in \mathcal{U}} J^u(x, t). \tag{4}$$

---

[2] The regulation includes limits on the duration of holdings and positions.

We say that a control $u^* \in \mathcal{U}$ is optimal if $J^{u^*}(x,t) = v(x,t)$ for all $x$ and $t$.

In general, finding an optimal control is difficult. However, a sufficient condition for optimality is given by the so-called *verification theorem* from the theory of controlled diffusions (see [Pha09]). Consider the infinitesimal generator $\mathcal{L}$ corresponding to the controlled diffusion above. Suppose that there exists a function $v \in C^{2,1}(\mathbb{R}^d \times [0,T])$ that solves the PDE,

$$\frac{\partial v(x,t)}{\partial t} + \mathcal{L}v(x,t) = -\min_{\alpha \in \mathbb{R}^d}\left\{\alpha^\top \nabla v(x,t) + c(x,\alpha)\right\}, \quad g(.,T) = C(.). \tag{5}$$

Then $v$ is the value function for (3) and the optimal control $v^*$ is given by,

$$u^*(x,t) = \arg\min_{\alpha \in \mathbb{R}^d}\left\{\alpha^\top \nabla v(x,t) + c(x,\alpha)\right\}. \tag{6}$$

The PDE in (5) is called the *Hamilton-Jacobi-Bellman* equation associated to the control problem (3).

## 2.2 Reinforcement Learning

In this section we provide a brief but self-contained overview of Reinforcement Learning (RL). We follow the exposition in [LKTF20] closely. RL is an approach to the problem of learning to control a dynamical system. It can be formulated via the general framework of Markov Decision Processes (MDP).

An MDP is given by $(\mathcal{S}, \mathcal{A}, \mathbb{P}, r)$, where $\mathcal{S}$ is a set of states, $\mathcal{A}$ is a set of actions (control values), $\mathbb{P}$ denotes the stochastic state transitions $\mathbb{P}(s_{t+1}|s_t, a_t)$ that describes the dynamics of the system, $\mathcal{R}$ denotes the reward: $r : \mathcal{S} \times \mathcal{A} \to \mathbb{R}$. The urgency in receiving the instantaneous reward in an MDP is encoded in the discount factor $\gamma \in (0,1]$.

In the RL approach to stochastic optimization, the goal is to learn a policy $\pi(a_t|s_t)$. In an MDP, a trajectory is a sequence of states and actions of length $T$, given by $\tau = (s_0, a_0, \ldots, s_T, a_T)$. The trajectory distribution $\mathbb{P}_\pi$ for the policy $\pi$ is given by

$$\mathbb{P}_\pi(\tau) = d(s_0)\prod_{t=0}^{T}\pi(a_t|s_t)\mathbb{P}(s_{t+1}|s_t, a_t),$$

where $d(s_0)$ is the distribution over the initial state.

The objective of the RL agent, $J(\pi)$, can then be written as an expectation under this trajectory distribution:

$$J(\pi) = \mathbb{E}_{\tau \sim \mathbb{P}_\pi}\left[\sum_{t=0}^{T}\gamma^t r(s_t, a_t)\right]. \tag{7}$$

In the following, we present a brief summary of solving RL problems. The high-level recipe for learning the agent is as follows: the agent *interacts* with a stochastic *environment* via a decision policy (control) $\pi(a|s)$, based on the current state $s_t$, selecting an action $a_t$, and then observing the resulting next state $s_{t+1}$ and reward value $r_t(s_t, a_t)$. By repeating these interactions, the *experiences* of transitions $\{(s_t^i, a_t^i, s_{t+1}^i, r_t^i)\}$ are used to update its policy.

**Policy gradients.** A direct approach to optimize the RL objective in Equation 7 is to parameterize the policy, $\pi = \pi_\theta(a_t|s_t)$ and calculate the gradients ($\theta$ might denote the weights of a deep network). One can express the gradient of the objective with respect to $\theta$ as:

$$\nabla_\theta J = \mathbb{E}_{\tau \sim \mathbb{P}_{\pi_\theta}(\tau)} \left[ \sum_{t=0}^{T} \gamma^t \nabla_\theta \log \pi_\theta(a_t|s_t) \underbrace{\left( \sum_{t'=t}^{T} \gamma^{t'-t} r(s_{t'}, a_{t'}) - b(s_t) \right)}_{\text{return estimate } \hat{A}(s_t, a_t)} \right], \quad (8)$$

where the return estimator $\hat{A}(s_t, a_t)$ can itself be learned separately (*critic* network). The baseline $b(s_t)$ can be estimated as the average reward over the sampled trajectories, or by using a value function estimator $V(s_t)$ (see below).

**Approximate dynamic programming** first estimates the state or state-action *value function* accurately, and then derives an optimal policy. A value function denotes the expected cumulative reward that will be obtained by following some policy $\pi(a_t|s_t)$ when starting from a given state $s_t$, in the case of the state-value function $V^\pi(s_t)$, or when starting from a state-action tuple $(s_t, a_t)$, in the case of the state-action value function $Q^\pi(s_t, a_t)$. We can define these value functions as:

$$V^\pi(s_t) = \mathbb{E}_{\tau \sim \mathbb{P}_\pi(\tau|s_t)} \left[ \sum_{t'=t}^{T} \gamma^{t'-t} r(s_t, a_t) \right]$$

$$Q^\pi(s_t, a_t) = \mathbb{E}_{\tau \sim \mathbb{P}_\pi(\tau|s_t, a_t)} \left[ \sum_{t'=t}^{T} \gamma^{t'-t} r(s_t, a_t) \right].$$

The fundamental recursive structure of these value functions is reflected in the *Bellman Backup* equation:

$$Q^\pi(s_t, a_t) = r(s_t, a_t) + \gamma \mathbb{E}_{s_{t+1} \sim \mathbb{P}(s_{t+1}|s_t, a_t)} \left[ V^\pi(s_{t+1}) \right], \quad (9)$$

or equivalently,

$$Q^\pi(s_t, a_t) = r(s_t, a_t) + \gamma \mathbb{E}_{\substack{s_{t+1} \sim \mathbb{P}(s_{t+1}|s_t, a_t) \\ a_{t+1} \sim \pi(a_{t+1}|s_{t+1})}} \left[ Q^\pi(s_{t+1}, a_{t+1}) \right]. \quad (10)$$

The Bellman backup equation (10) is the basis of two commonly used dynamic programming algorithms in RL: Q-learning and actor-critic methods. In Q-learning, the optimal policy is derived *greedily* from the Q-function, as $\pi^*(a_t|s_t) = \delta(a_t = \mathrm{argmax} Q(s_t, a_t))$, hence the goal turns into finding the optimal Q-function via the condition:

$$Q^*(s_t, a_t) = r(s_t, a_t) + \gamma \mathbb{E}_{s_{t+1} \sim \mathbb{P}(s_{t+1}|s_t, a_t)} \left[ \max_{a_{t+1}} Q^*(s_{t+1}, a_{t+1}) \right]. \quad (11)$$

To cast this equation into a learning algorithm, one minimizes the difference between the left-hand side and right-hand side of this equation with respect to the parameters of a parametric Q-function estimator with parameters $\phi$, $Q_\phi(s_t, a_t)$. Among the many variations of the original *Q-learning* is a variant used in deep reinforcement learning which employs a replay buffer and taking gradient steps on the Bellman error (the square of the difference between left and right hand side of (11)) while interacting with the environment and collecting data. Moreover, in such implementations of Q-learning, a *target network* is involved, where the target value $r_i + \gamma \max_{a'} Q_{\phi_k}(s', )$ is parametrized by the weights $\phi_L$, where $L$ is a lagged iteration.

**Actor-critic methods.** While the policy gradient formula in equation (8) provides a direct way of optimizing the policy based on the observed roll outs, it suffers from *high variance* originating from the to the stochastic nature of the environment and the variance of the policy.

One way to reduce this variance is through *advantage functions*. The advantage function $A(s_t, a_t)$ represents how much better it is to take a specific action $a_t$ at state $s_t$, compared to the average quality of actions following the same policy:

$$A(s_t, a_t) = Q(s_t, a_t) - V(s_t), \tag{12}$$

where $V(s_t)$ is the value function.

Combining policy gradients with advantage functions has given rise to a rich family of RL training algorithms. Actor-Critic (AC) algorithms are a combination of policy gradients and approximate dynamic programming. They involve *both* a parameterized policy and a parameterized value function. The value function is used to provide a better estimate of the advantage function, $\hat{A}(s, a)$, for policy gradient calculation. In the Q-learning approach, one tries to directly learn the optimal Q-function. In contrast, the actor-critic method learns the Q-function corresponding to the current parameterized policy $\pi_\theta(a|s)$, satisfying,

$$Q^\pi(s_t, a_t) = r(s_t, a_t) + \gamma \mathbb{E}_{\substack{s_{t+1} \sim \mathbb{P}(s_{t+1}|s_t, a_t) \\ a_{t+1} \sim \pi_\theta(a_{t+1}|s_{t+1})}} \left[ Q^\pi(s_{t+1}, a_{t+1}) \right].$$

Actor-critic algorithms leverage a policy evaluation and a policy update. The policy evaluation phase computes the Q-function for the current policy $\pi$, $Q^\pi$, by minimizing the Bellman error. In the next step, the policy update is done by computing the policy gradient (see [LKTF20] for more details).

**Advanced Actor-Critic: DDPG, TRPO, PPO** Proximal Policy Optimization (PPO) aims to control the policy update by penalizing "too much distance"[3].

PPO is related to a parent RL training algorithm called the *Trust Region Policy Optimization* (TRPO). TRPO controls the change in (probabilistic) policies by bounding the Kullback-Leibler (KL) divergence between the policies:

$$\begin{aligned} \text{maximize}_\theta \ & \hat{\mathbb{E}}_t \left[ \frac{\pi_\theta(a_t|s_t)}{\pi_{\theta_\text{old}}(a_t|s_t)} \hat{A}_t \right], \\ \text{subject to} \ & \hat{\mathbb{E}}_t \left[ \text{KL}(\pi_{\theta_\text{old}}(\cdot|s_t), \pi_\theta(\cdot|s_t)) \right] \leq \delta, \end{aligned} \tag{13}$$

where $\theta_\text{old}$ is the old policy parameters before the update.

---

[3] drastic change in policy performance between the two consecutive updates, sometimes called "falling off the cliff"

The PPO algorithm tries to control the change in the policies by penalizing the KL divergence:

$$\mathcal{L}_{\text{ppo-penalty}}(\theta) = \hat{\mathbb{E}}_t \left[ \frac{\pi_\theta(a_t|s_t)}{\pi_{\theta_{\text{old}}}(a_t|s_t)} \hat{A}_t \right] - \beta \text{KL}(\pi_{\theta_{\text{old}}}(\cdot|s_t), \pi_\theta(\cdot|s_t)), \quad (14)$$

where $\beta$ is the penalty hyper-parameter, we call this version of the algorithm, PPO-penalty.

Another variation of the PPO algorithm uses a "clipped" policy:

$$\mathcal{L}_{\text{ppo-clip}}(\theta) = \hat{\mathbb{E}}_t \left[ \min \left( \frac{\pi_\theta(a_t|s_t)}{\pi_{\theta_{\text{old}}}(a_t|s_t)} \hat{A}_t, \text{clip}\left( \frac{\pi_\theta(a_t|s_t)}{\pi_{\theta_{\text{old}}}(a_t|s_t)}, 1 - \epsilon, 1 + \epsilon \right) \hat{A}_t \right) \right]. \quad (15)$$

The clip function limits the value of policy ratio between $(1 - \epsilon, 1 + \epsilon)$.

## 3 Fair Transfer Price á la Bergault & Gueant

In [BG23], the problem of corporate bond market making when the securities are illiquid is considered. They explore whether the market maker can use other sources of information for price discovery, when liquidity is scarce. For a market maker with a large enough market share, they note that the intensity of bid-ask RFQs received is an important source of information:

- The direction of RFQs (buy/sell) indicates the sentiment of the clients and could be used, in aggregate as a proxy for the market view towards the given security,

- The clients answer to the RFQ response (trade/trade away or cancel) inform about the demand curve of clients.

For modelling the variability in liquidity, they assume that the market maker receives RFQs which are governed by a Poisson process where the intensities are also stochastic. More concretely, the bid-ask RFQ flow is modeled by a two dimensional Markov-modulated Poisson process (MMPP) which we describe below.

The number of RFQs received by the market maker for each bond at the bid and at the ask is modelled as a two point Poisson process. It is assumed that intensities $\lambda_t^a$ (for ask-RFQ) and $\lambda_t^b$ (for bid RFQ) are continuous-time Markov chains with values in a finite set. To take into account asymmetries in liquidity, we assume that the intensity process $\lambda_t = (\lambda_t^b, \lambda_t^a)$ takes values in $\{\lambda^{1,b}, \cdots, \lambda^{m_b,b}\} \times \{\lambda^{1,a}, \cdots, \lambda^{m_a,a}\}$ with a transition matrix $Q \in \mathbb{M}_{m_a \times m_b}$.

In this framework, the dynamics of the bond price are given by arithmetic Brownian motion:

$$dS_t = \kappa(\lambda_t^a - \lambda_t^b)dt + \sigma dW_t. \quad (16)$$

The trader answers to bid-ask RFQs by providing the quotes $S_t - \delta_t^b$ and $S_t + \delta_t^a$ respectively. We assume that the probability of success in RFQ response is calibrated in sigmoid-type functions $f^b(\delta_t^b)$ and $f^a(\delta^a_t)$ (see figure 1). Assuming that the RFQ sizes are a fixed lot $z$, the inventory of the market maker changes as follows,

$$dq_t = z(dN^b_t - dN^a_t). \tag{17}$$

As the RFQs flow in, the cash flow evolves based on,

$$dX_t = zS^a_t dN^a_t - zS^b_t dN^b_t, \tag{18}$$

and the desk P&L is given by P&L$_t = (X_t + q_t S_t)$. The main construct in the definition of FTP is the objective of the trader, which is defined as acting based on an *RFQ quote strategy* $(\delta^b_t, \delta^a_t)$ to maximize the objective,

$$J_{(\delta^b_t, \delta^a_t)}(q) = \mathbb{E}\left[\int_0^T dP\&L_t - \frac{\lambda}{2} d\langle P\&L \rangle_t\right], \tag{19}$$

where $T$ is a fixed time horizon and $\langle , \rangle$ denotes the quadratic variation. Assuming that the trader is able to identify the state of the RFQ intensities $(\lambda^{j_b,b}, \lambda^{j_a,a})$ is at any point in time, the authors write down and solve (by approximation of the value functions) the resulting system of HJB equations for the family of value functions $\{\theta^{j_a,j_b}\}_{1 \leq j_b \leq m_b, 1 \leq j_a \leq m_a}$. At this point, the conceptual argument provided by the authors is the following:

*"If trading flows (or intensities in mathematical models) at the bid and at the ask are the same, then the optimal bid and ask prices of a market maker with no inventory will be symmetric around the reference price, which is therefore a fair transfer price. What happens, however, when a market maker is aware of asymmetries in the trading flows is that they skew their quotes even in the absence of inventory. As a consequence, the average between the optimal bid and ask quotes ceases to coincide with the reference price. However, it is a fair transfer price given the current context in terms of liquidity."*

With this, finally, denoting the optimal value function by $\theta^{j_b,j_a}$, the following steps are taken to define the FTP:

- An infinite horizon limit of the value function is shown to exist by assuming the intensity process has certain property:

$$\lim_{T \to \infty} \theta^{j_b,j_a}(t,q) - c(T-t) = \theta^{j_b,j_a}_\infty(q), \tag{20}$$

  for some constant c.

- The skew effect of market liquidity on the price is defined as,

$$\text{skew}^{j_b,j_a}_\infty = \bar{\delta}^a \left(\frac{\theta^{j_b,j_a}_\infty(0) - \theta^{j_b,j_a}_\infty(-z)}{z}\right) - \bar{\delta}^b \left(\frac{\theta^{j_b,j_a}_\infty(0) - \theta^{j_b,j_a}_\infty(z)}{z}\right). \tag{21}$$

- The FTP is defined as the average between the ask-RFQ and bid-RFQ quotes by a trader with no inventory and infinite time horizon (see [BG23] for more details),

$$S^{\text{FTP}}_t = S_t + \frac{1}{2}\text{skew}_\infty. \tag{22}$$

# 4  A Finite Horizon RFQ Quoting Strategy

While the FTP framework possesses favorable properties as a quoted price of illiquid OTC securities (see Section 3.2.2 in [BG23] for a discussion), from a practitioner point of view, we observe some potential restrictions:

- The pricing methodology is based on an elaborate calibration methodology for estimating the trade flow (RFQ intensities). However, estimating the current state of the intensities is difficult.

- The underlying dynamics of the market price is modelled by an arithmetic Brownian motion. While the authors show reasonable behaviour of FTP under constant shift of the dynamics, it is rather unclear how the model behaves when deployed in practice, where deviations from the written dynamics model is likely.

- A main component of the FTP is the existence of the *infinite time horizon* limit for the value function (the optimal objective). While this is justified mathematically, it may be rather hard to digest from the practitioner point of view- as the trader usually engages with an RFQ, based on an existing fixed time horizon for achieving the firm/desk objectives. It should also be noted that regulation requires a bond trading desk in a bank to hold the bonds in the inventory for only a limited time period.

With the above points in mind, we propose a simplification of the FTP optimization objective and consider the trader responding to the RFQs trying to optimize the objective (19) within a pre-determined time horizon. We fix a finite trading time horizon $T$ and consider the discretization $0 = t_0 < t_1 < \cdots t_p = T$, the objective of the bond trader is to provide optimal bid and ask-RFQ quotes via bid-ask spreads $(\delta_t^b, \delta_t^b)$ to maximize the expected P&L while penalizing the path dependent variability based on equation (19).

When discretized, the trader's objective takes the form,

$$J^{\text{discretized}}_{(\delta_t^b, \delta_t^a)}(q) = \mathbb{E}\left[\sum_{i=1}^{p} \Delta \text{P\&L}_{t_i} - \frac{\lambda}{2} \sum_{i=1}^{p} \Delta \text{P\&L}_{t_i}^2\right]$$
$$= \mathbb{E}\left[\sum_{i=1}^{p} \left(\Delta \text{P\&L}_{t_i} - \frac{\lambda}{2}(\Delta \text{P\&L}_{t_i})^2\right)\right] \quad (23)$$

The expression in the summation above can be thought as the step-wise reward for an RL agent, tasked with estimating the optimal response to the incoming bid-ask RFQs within a fixed time horizon. In the next section we explain, the deep reinforcement learning approach for parameterizing and training the agent.

## 4.1 Experiments

In this section, we demonstrate how the optimal RFQ response strategy could be estimated by data-driven decision policy learning methods such as RL.

**Data Simulation** We skip the parameter estimation and replicate the exact simulation environment, based on the specified model and parameters calibrated in [BG23][4]. We assume that both bid and ask RFQ intensity processes take value in a set $\{\lambda_1 = 10.83, \lambda_2 = 73.03\}$[5]. We set the generator matrix for the MMPP as:

$$Q = \begin{pmatrix} -14.01 & 4.37 & 4.37 & 5.27 \\ 19.32 & -60.91 & 12.54 & 29.05 \\ 19.32 & 12.54 & -60.91 & 29.05 \\ 23.67 & 15.00 & 15.00 & -53.67 \end{pmatrix} \qquad (24)$$

The probability of bid, ask-RFQ quotes are represented by a single S-curve:

$$f^b(\delta) = f^a(\delta) = \frac{1}{1 + \exp\left(\alpha + \frac{\beta}{\delta_0}\delta\right)}, \qquad (25)$$

where $\delta_0 = 0.09$ is the current composite bid-ask spread of the bond, and the parameters $\alpha = -0.7$ and $\beta = 3.1$ are estimated with a logistic regression (Figure 1 below).

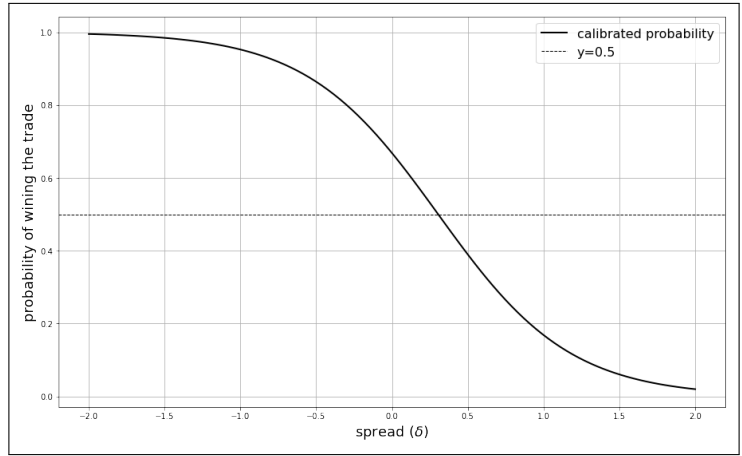

Figure 1: The calibrated S-curve for the RFQ success, based on the magnitude of the spread.

To simulate the price process based on (3), we set $\kappa = 2.99$ and the volatility as $\sigma = 18.39$ (annualized), with the initial mid-price at $S_0 = 103.593$ (see figures 2, 3 below).

---

[4]The parameters are calibrated to the daily bid and ask-RFQ counts, based on some proprietary data the authors had access to.

[5]The intensities and the Q-matrix values are taken from Table 1, page 20 in [BG23]

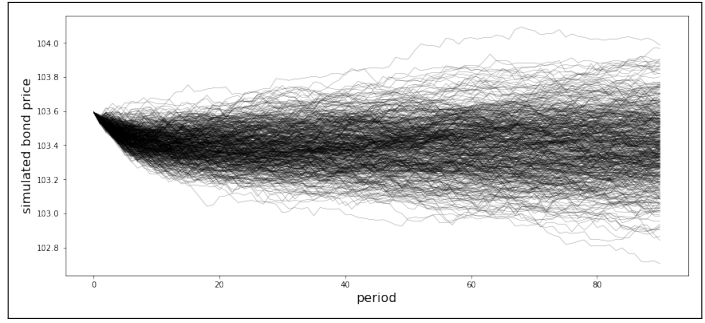

Figure 2: Simulated bond price. Note that the visible downward sloping movements are due to our chosen intensity parameters for the incoming bid-ask RFQs (see equation 3).

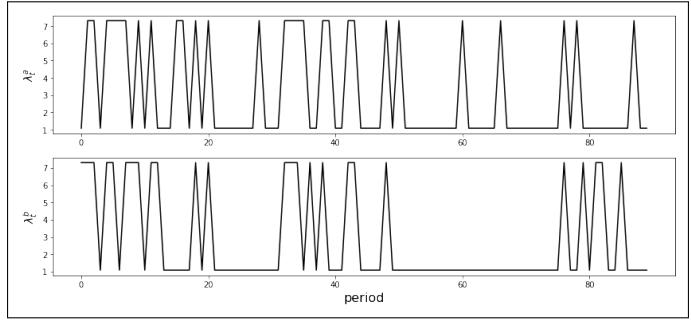

Figure 3: A sample path of the simulated (bid, ask) RFQ intensity process

## 5 RL Agent Configuration

**The Simulation Sampler:** All RL agents that we implemented rely on the sampler. The inputs to the sampler are the $Q$ matrix, the $\lambda_{low} = 10.83$ and $\lambda_{high} = 73.03$ values, the number of simulation days ($N$), the initial price, initial $\lambda$ state, the $\sigma$ value, and the $\kappa$ value. Using these input values, the sampler simulates a Markov process with 4 states. These states are the four possible $(\lambda^b, \lambda^a)$ combinations: low-low, low-high, high-low, and high-high. Using these $\lambda$ states and the provided parameters, the sampler simulates the price process. The sampler yields three matrices of shape $B \times N$ to the RL agent implementation: the $\lambda^b$ matrix, the $\lambda^a$ matrix, and the price matrix ($B$ is the batch size).

### 5.1 The Environment

Valid RL environments stipulate the observation space, the action space, the reward function, and structure to inform the agent when an episode is complete. We simulate episodes of the length 30-days.

**The State Space:** Because we assume that the agent is a market-maker, we provide it with the current $\lambda^b$ and $\lambda^a$ values. The other two inputs to the agent are the current time represented as a fraction of the total time horizon, and the current inventory level.

**The Action Space:** The agent must choose two actions at each time step, $\delta^b$ and $\delta^a$. These controls dictate the price that the agent is quoting for bid RFQs and ask RFQs, respectively: $S_t^b = S_t - \delta^b$ and $S_t^a = S_t + \delta^a$. Practically, these spreads can be any value, however, continuous-action RL algorithms benefit from more confined action spaces. Further, PPO and DDPG algorithms benefit from continuous action spaces that are symmetric.

For symmetry, we define the action space for each $\delta$ as continuous values ranging from −1 and 1. Then, we transform these values using min-max scaling to range from −0.16 to 0.2. A $\delta$ value of −0.16 gives a 99.8% chance of losing each RFQ and a value of 0.2 gives a 99.8% chance of winning each RFQ (see (25)). By re-scaling the input in this way, we further confine the action space to expedite training convergence.

**The Reward Function:** Internally, the environment is computing the agent's cash process, step-wise P&L and inventory value process as the agent captures the bid-ask spread. For each episode, the cumulative reward is calculated based on the sum inside the expectation in equation (23).

## 5.2 Training the RL Agents

The two RL algorithms we utilized in this paper are PPO & DDPG. We trained the RL agents in Python using the Stable Baselines-3 library [RHG+21]. To speed up training time, we wrapped our RL environment implementation in the Stable Baselines-3 `VecEnv` wrapper to train on multiple environments in parallel.

**PPO:** Because the PPO algorithm is sample inefficient, we ran this algorithm using 512 environments in parallel. Overall, we found the convergence of this algorithm to occur naturally with little parameter tuning. The parameters we used to train the agent using the PPO algorithm are listed in Table 1. A sample cumulative reward curve is shown in Figure 4.

**DDPG:** The DDPG algorithm required significantly more training time. For exploration noise, we tried both dynamic noise based on an Ornstein Uhlenbeck process¢ and normal action noise, but ultimately found that the DDPG algorithm failed to converge in most instances (see Table 2 for the chosen parameters).

**PPO Network Architecture:** With respect to the PPO algorithm, we used a fully-connected MLP with two hidden layers of 64 units and tanh activation functions. This is the same network configuration as the one detailed in the original PPO paper [SWD+17].

| Parameter | Value |
|---|---|
| learning_rate | 0.0003 |
| $\gamma$ (discount factor) | 0.99 |
| GAE_lambda | 0.95 |
| Policy clip $\epsilon$ | 0.2 |

Table 1: PPO RL agent training parameters

| Parameter | Value |
|---|---|
| learning_rate | 0.001 |
| $\gamma$ (discount factor) | 0.99 |
| GAE_tau | 0.005 |
| buffer_size | 1000000 |
| action_noise | $\mathcal{N}(\mu = 0, \sigma = 0.6)$ |

Table 2: DDPG RL agent training parameters (standard parameters in Stable Baseline-3)

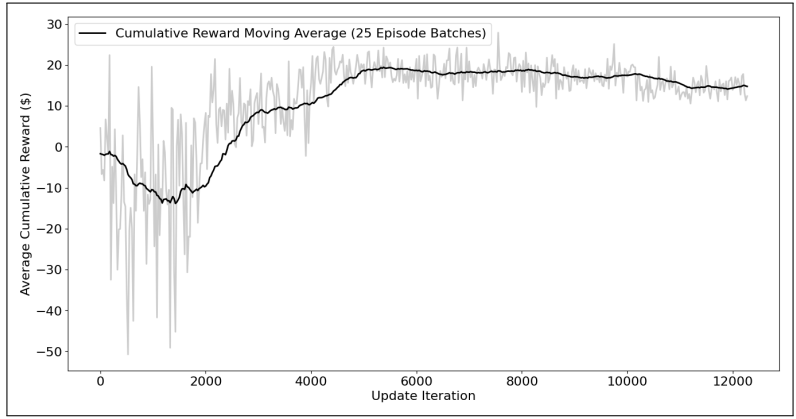

Figure 4: A sample reward curve generated by training a PPO RL agent on the default Q, $\lambda$, and $q_{init}$ parameters. The agent converges to an average cumulative reward of about $20.

## 6 Outcome Analysis

**Baseline**   Using the "Baseline" parameters stipulated in Table 3, we trained a PPO RL agent using the parameters from Table 1 until convergence. Figure 5 illustrates the cumulative reward of the agent at each time-step over the time horizon for the trained agent.

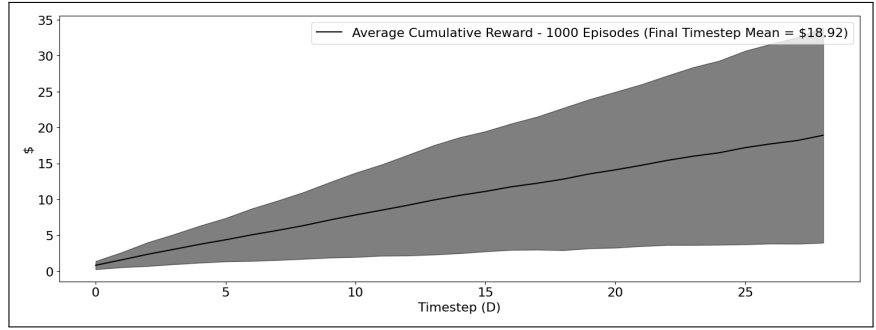

Figure 5: The average cumulative reward of the agent at each time-step trained using the "Baseline" parameters from Table 3. The agent achieves a final time-step average cumulative reward over 1,000 simulations of $18.92.

Figure 6 shows the $\kappa$-implied price movement and the $\delta$ values yielded by the agent at each time-step for a single training episode. It can be seen from this plot (and from Figure 7, which shows the average $\delta^b$ and $\delta^a$ values over the time horizon), that the agent has learned to yield positive spread values in the $(0, 0.1)$ range. Further, when the $\kappa$-implied price movement is positive, meaning the price will move upwards in the near future, the agent reduces $\delta^b$ to purchase more of the bond and vice-versa.

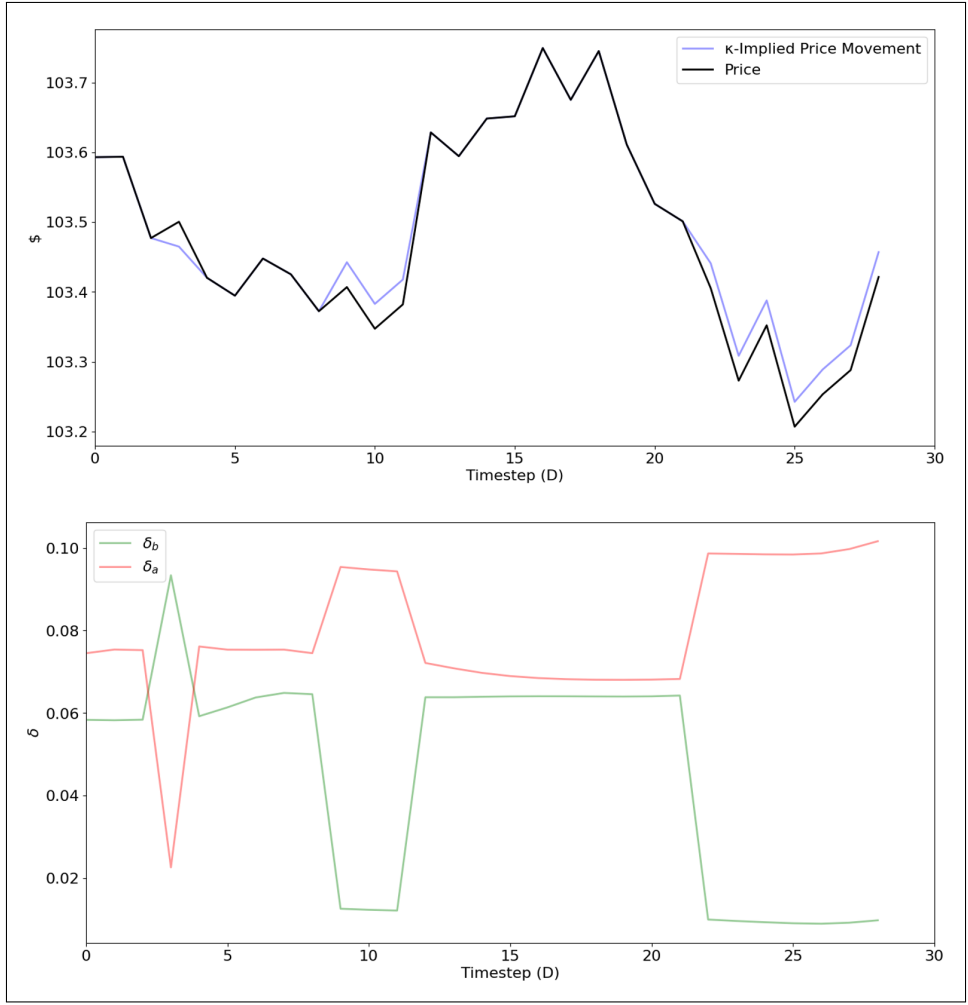

Figure 6: The top plot delineates the $\kappa$-implied price movement relative to the current price. The $\kappa$-implied price movement is the anticipated price movement as a function of the current $\lambda^a$ and $\lambda^b$ values. Mathematically, it is defined as $S_t + \kappa \cdot (\lambda^a - \lambda^b)$. The visualization is useful as it shows which direction price will go next. The bottom plot illustrates the $\delta^b$ and $\delta^a$ values selected by the agent at each time-step. As expected, the agent chooses these values responsively according to the privileged $\lambda^b$ and $\lambda^a$ information.

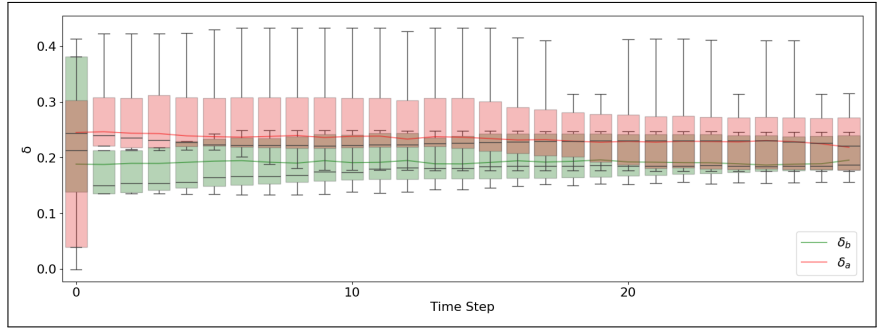

Figure 7: This figure shows the average $\delta^b$ and $\delta^a$ values over the time horizon for the "Baseline" parameter setting described in Table 3

**The Impact of the Initial $\lambda$ State**  Figure 8 shows the average instantaneous price as a function of the time-step for the "Baseline" parameters in Table 3. Future price is normally distributed with mean 0. Because the $Q$ matrix in Eq. 24 is symmetric about the $\lambda^a = \lambda_{low}, \lambda^b = \lambda_{high}$ and $\lambda^a = \lambda_{high}, \lambda^b = \lambda_{low}$ states, and the initial $\lambda$ state is random, we expect this to be the case.

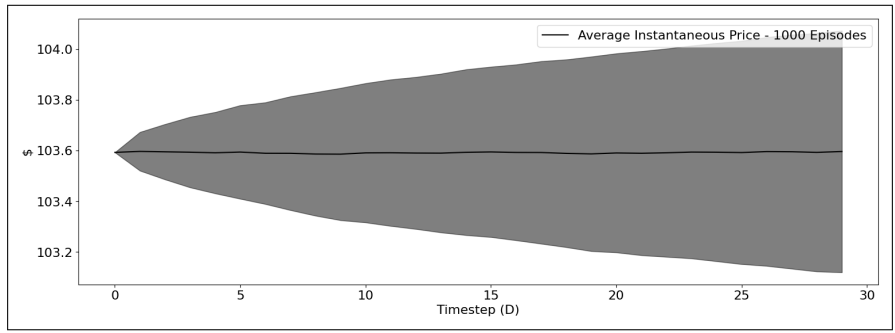

Figure 8: The average instantaneous price as a function of the timestep using a 1,000 sample batch. The grey area marks the standard deviation of the distribution.

In this section, we attempt to visualize the impact of the initial $\lambda$ state on the agent's behavior. Using the "Neg. Init. Price Trajectory" and "Pos. Init. Price Trajectory" parameters in Table 3, we observe the price processes in figures 9 and 10.

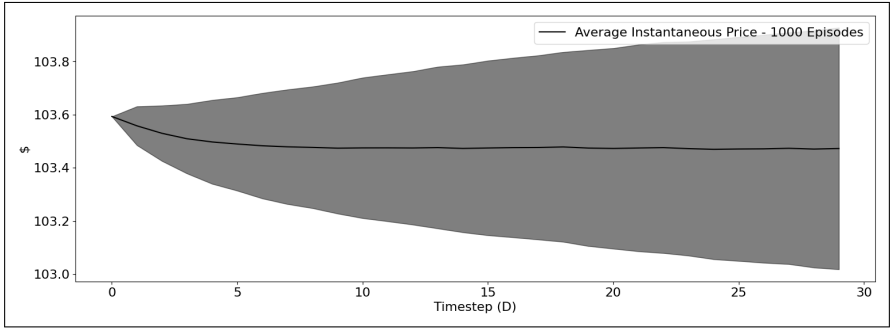

Figure 9: The average instantaneous price when the initial $\lambda$ state is $\lambda_a = \lambda_{low}, \lambda_b = \lambda_{high}$.

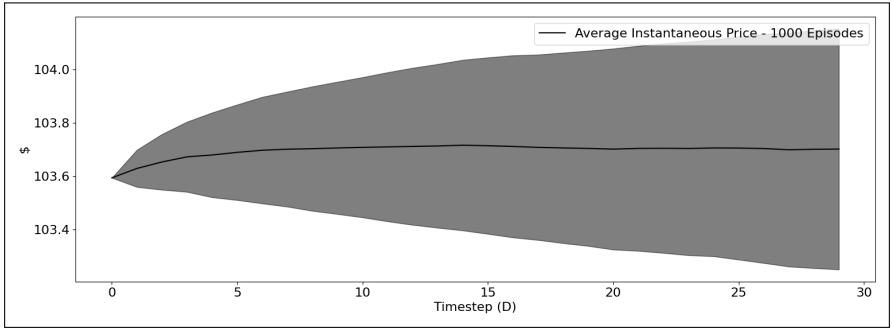

Figure 10: The average instantaneous price when the initial $\lambda$ state is $\lambda_a = \lambda_{high}, \lambda_b = \lambda_{low}$.

Table 3: Parameter values for the baseline experiment, and the negative & positive initial price bias experiments.

| Parameter | Baseline | Neg. Init. Price Trajectory | Pos. Init. Price Trajectory |
| --- | --- | --- | --- |
| $\lambda_{low}$ | | 10.83 | |
| $\lambda_{high}$ | | 73.03 | |
| Q | | Eq. 24 | |
| # simulation days | | 30 | |
| Initial price | | 103.593 | |
| Initial $\lambda$ state | Random | $\lambda_a = \lambda_{low}, \lambda_b = \lambda_{high}$ | $\lambda_a = \lambda_{high}, \lambda_b = \lambda_{low}$ |
| Initial inventory | | 0 | |
| $\sigma$ | | 18.39 | |
| $\kappa$ | | 2.29 | |
| $z$ | | 1 | |

Given these price processes, we expect the agent to initially favor a inventory process in the direction of the initial price move. Afterwards, as the price process smooths out, we expect the agent to revert to a similar strategy to the baseline strategy illustrated in figures 6 and 7, in which the agent provides similar quotes for $\delta^b$ and $\delta^a$, and we expect the agent to react to the $\lambda^b$ and $\lambda^a$ data responsively.

Figures 11 and 12 and figures 13 and 14 illuminate the behavior of the trained RL agent using a negative initial price setting and a positive initial price setting, respectively. These figures align with our original hypothesis. For the negative initial price parameter setting, the agent initially discounts the ask quotes and aggressively attempts to short the bond. As time progresses, the agent reverts back to the behavior we saw in Figure 7, in which similar quotes are provided near a $\delta$ value of 0.25 for $\delta^b$ and $\delta^a$. In the positive initial price parameter setting, the behavior is reversed. The agent aggressively tries to buy the bond because it is anticipating an upwards move in the bond price. Then, as time progresses, reverts back to a responsive quoting strategy averaging around a $\delta$ value of 0.25.

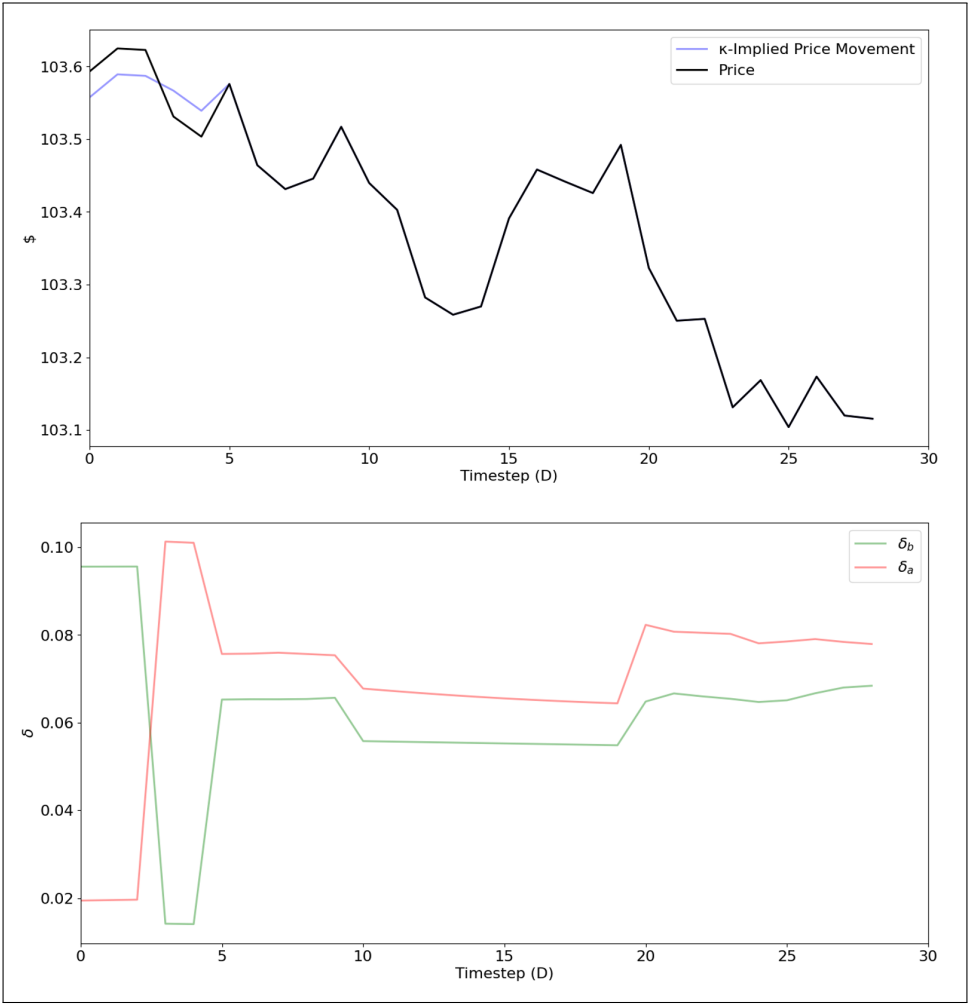

Figure 11: The $\kappa$-implied movement using the negative initial price parameters from Table 3.

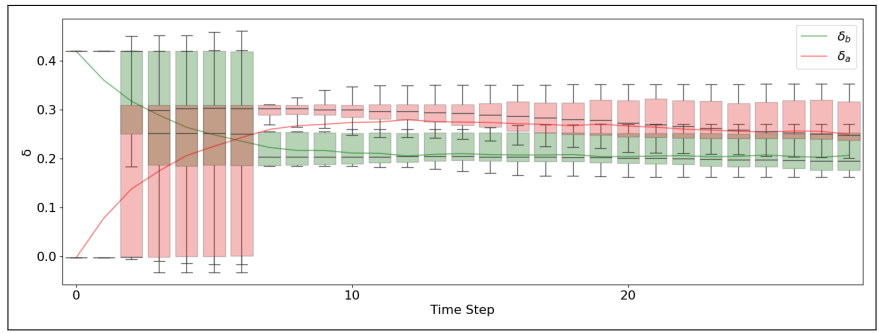

Figure 12: A box-plot showing the distribution for $\delta_b$ and $\delta_a$ actions over the time horizon using the positive initial price parameters from Table 3.

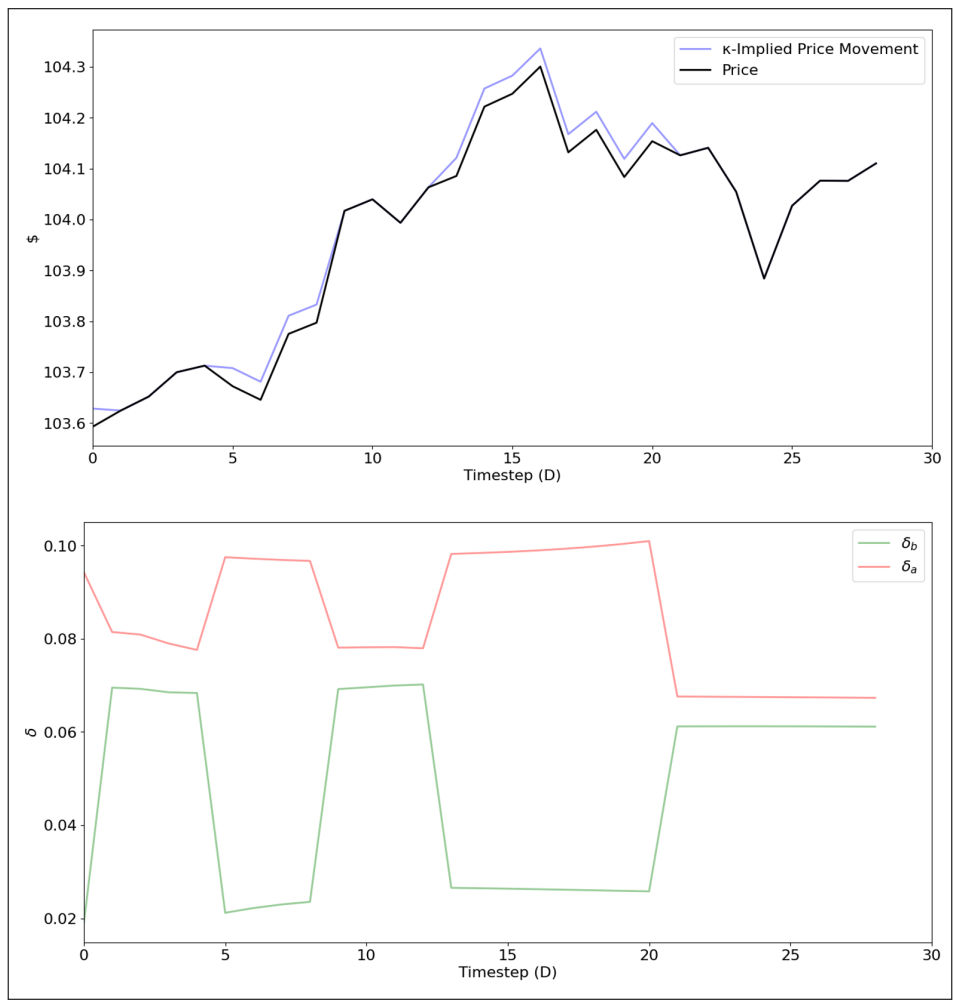

Figure 13: The $\kappa$-implied movement using the positive initial price parameters from Table 3.

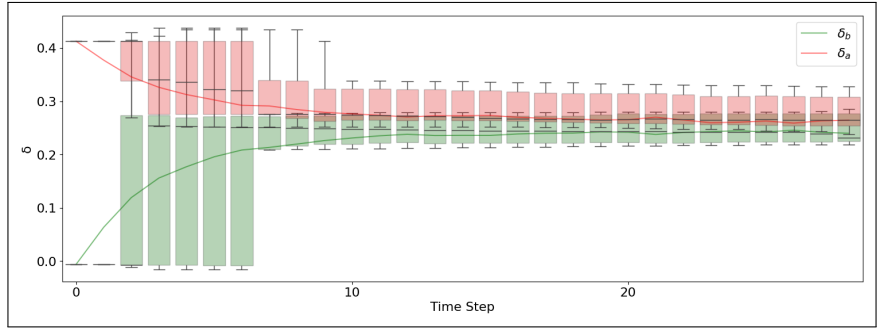

Figure 14: A box-plot showing the distribution for $\delta^b$ and $\delta^a$ actions over the time horizon using the negative initial price parameters from Table 3.

This analysis of the initial price state visualizes the agent's ability to react to short-term insights into the direction of the price movement. To simulate this short-term price movement, we modified the initial $\lambda$ state to favor an initial jump or drop in price. To simulate a longer-term drift in the bond's price, we can modify the values of the $Q$ matrix in Eq. 24. Because the $Q$ matrix from Eq. 24 is symmetrical about the $\lambda^a = \lambda^{low}, \lambda^b = \lambda^{high}$ and $\lambda^a = \lambda^{high}, \lambda^b = \lambda^{low}$ states, the probability that we will be in either one of these states given a random initial state is equivalent.

More specifically, the stationary state distribution of the $Q$ matrix from Eq. 24 is [0.602, 0.109, 0.109, 0.178], meaning there is a 10.9% chance that we will be in the low-high and high-low $\lambda^a$-$\lambda^b$ states as $t \to \infty$. The $Q$ matrix given by Eq. 26 yields a stationary state of [0.511, 0.200, 0.105, 0.182], meaning there is now a 20% chance that there will be in a low-high $\lambda^a$-$\lambda^b$ state and a 10.5% chance that there will be a high-low $\lambda^a$-$\lambda^b$ state as $t \to \infty$. Because the probability distribution favors a low-high $\lambda^a$-$\lambda^b$ state, price will naturally drift downwards over time. We can simulate the opposite scenario using the $Q$ matrix in Eq. 27, which yields a stationary state of [0.511, 0.105, 0.200, 0.182]. Simulated price processes for the $Q$ matrices from Eq. 26 and Eq. 27 are shown in figures 15 and 16, respectively. From these figures, we can see the long-term price drift stipulated by the imbalanced stationary-state distributions previously mentioned.

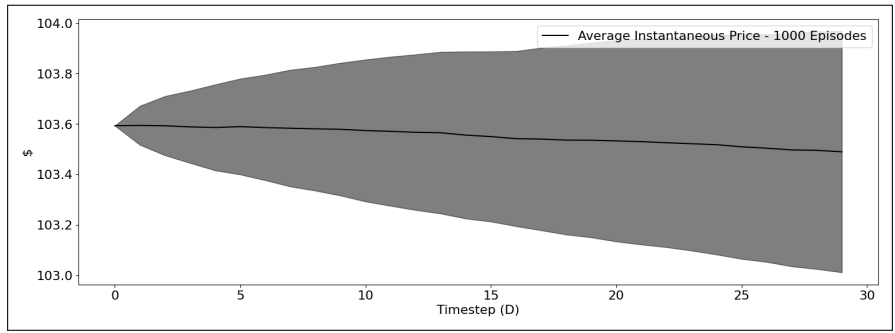

Figure 15: The price distribution over 1,000 episodes using the $Q$ matrix with a negative bias from Eq. 26.

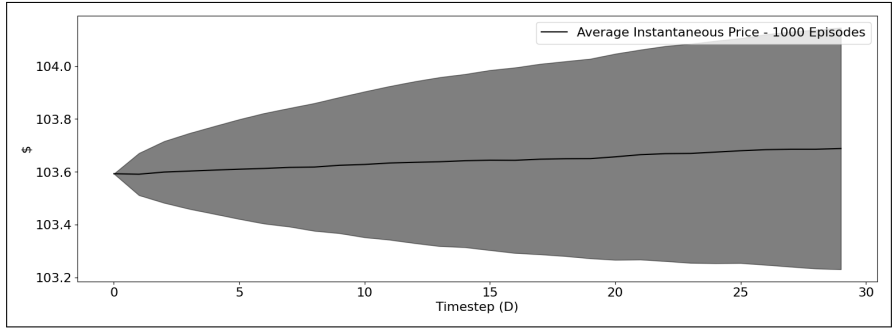

Figure 16: The price distribution over 1,000 episodes using the $Q$ matrix with a positive bias from Eq. 27.

$$Q = \begin{pmatrix} -20.01 & 10.37 & 4.37 & 5.27 \\ 19.32 & -60.91 & 12.54 & 29.05 \\ 19.32 & 22.54 & -70.91 & 29.05 \\ 23.67 & 25.00 & 15.00 & -63.67 \end{pmatrix} \quad (26)$$

$$Q = \begin{pmatrix} -20.01 & 10.37 & 4.37 & 5.27 \\ 19.32 & -70.91 & 22.54 & 29.05 \\ 19.32 & 12.54 & -60.91 & 29.05 \\ 23.67 & 25.00 & 15.00 & -63.67 \end{pmatrix} \quad (27)$$

Because the $Q$ matrices from Eq. 26 and Eq. 27 are symmetrical about the low-high and high-low $\lambda^a$-$\lambda^b$ states, we expect the behavior of agents separately trained on these $Q$ matrices to be symmetrical as well. Further, we expect the behavior of an agent trained on the $Q$ matrix with negative long-term bias to favor selling and to prefer buying when trained on the $Q$ matrix with positive long-term bias. Figures 17 and 18 illustrate the behavior of agents trained on the $Q$ matrix with negative and positive biases, respectively. As expected, these $\delta$ box-plots are flipped versions of each other. The agent provides consistent discounts on sell RFQs for the negatively biased $Q$ matrix (as shown in Figure 17), and discounts on buy RFQs for the positively biased $Q$ matrix from Figure 18. These strategies yield average quantity processes illustrated in figures 19 and 20, respectively. To no surprise, these strategies yield quantity processes that gradually decrease and increase for the agents trained on the negatively and positively-skewed $Q$ matrices described above.

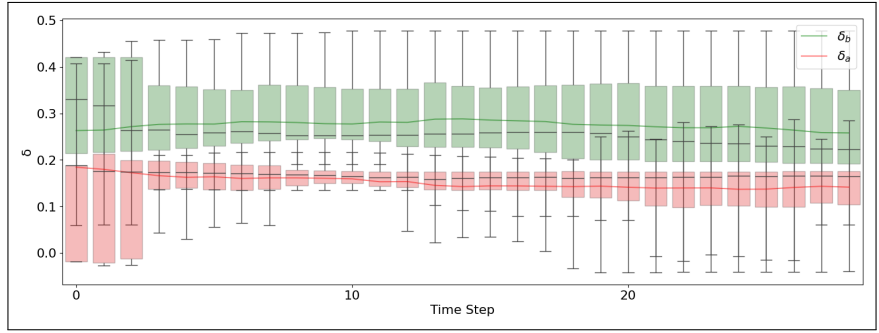

Figure 17: A box-plot showing the distribution for $\delta_b$ and $\delta_a$ actions over the time horizon using the negatively-skewed $Q$ matrix from Eq. 26.

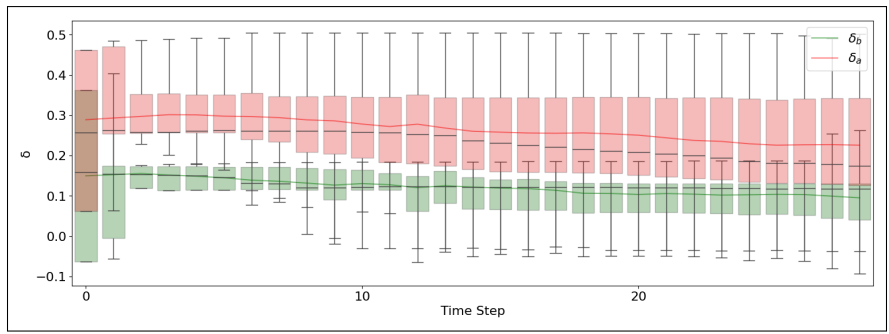

Figure 18: A box-plot showing the distribution for $\delta_b$ and $\delta_a$ actions over the time horizon using the positively-skewed $Q$ matrix from Eq. 27.

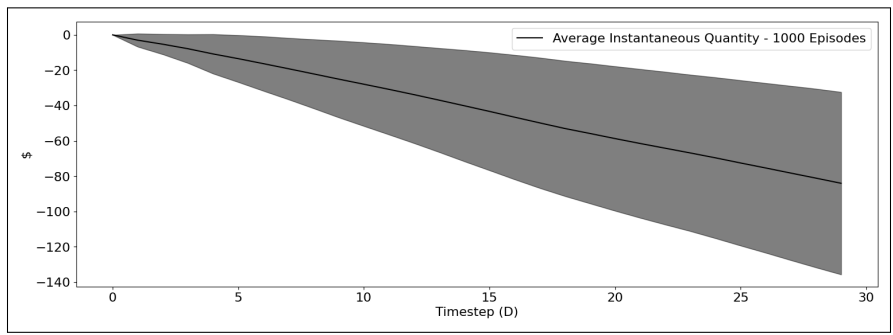

Figure 19: A simulation of the agent's quantity process when trained on the negatively-skewed $Q$ matrix from Eq. 26 over the time horizon a 1,000 episode sample.

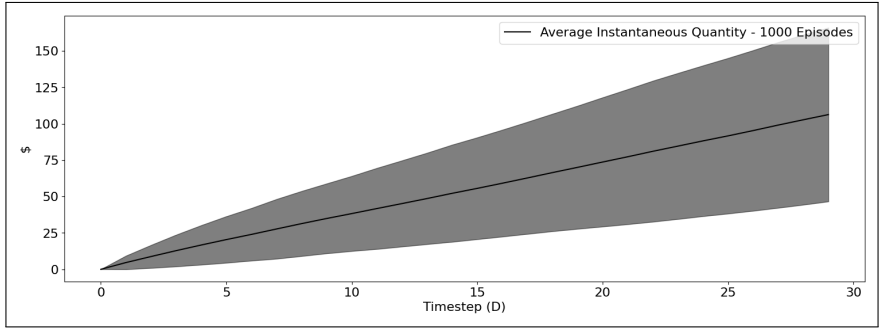

Figure 20: A simulation of the agent's quantity process when trained on the positively-skewed $Q$ matrix from Eq. 27 over the time horizon a 1,000 episode sample.

Another interesting behavior of the agents is that they do not refuse to buy bonds altogether when the price has a negative drift, nor do they refuse to sell bonds when the price has a positive drift. Instead, the agents slightly skew their quotes to factor in the long-term price movement of the bond. Further, unlike in the previous experiment, in which short-term price movement information was available to the agents, the trained agents did not *aggressively* short or buy in the direction of the price-skew in the first time-step. Their positions were obtained gradually to capture a portion of the well-priced bid/ask RFQs in the opposite direction of the anticipated price movement.

# Conclusion & Future Directions

In this paper, we considered the problem of training an autonomous RL agent which follows the optimal bid-ask RFQ quoting strategy in a data driven manner. Inspired by the Fair Transfer Pricing framework introduced by Bergault and Gueant, we chose the finite time horizon version of the optimal control problem in [BG23] as the RL learning objective. We demonstrated that the PPO algorithm for training the RL agent results in a stable training convergence. We also performed outcome analysis to examine the reasonableness of the trained agent's behavior.

We trained the RL algorithm using the training episodes simulated based on a model for the price dynamics and a model for the bid-ask RFQ flow. Examination of the performance of the model when transferred from simulation to the real world is crucial[6]. The authors note that this aspect of deploying adaptive autonomous agents in real world trading scenarios is widely understudied and perhaps points to a gap hindering wider industry adoption of AI-based autonomy in capital markets.

As mentioned in the introduction, one can also take an *offline RL* approach and train the agent on the desk's RFQ historical data without any need to a simulation model. Along these lines, applications of the recent paper [STZnt] introducing a mixture of both model free and model-based approaches for pricing is an interesting direction.

---

[6] Sim2Real Transfer risk is an active area of research in other domains where RL agents are being deployed in safety critical scenarios.

**Acknowledgment** Authors thank Philippe Bergault for helpful discussions about the FTP framework and calibration of the model parameters in [BG23]. We also thank Matt Davison for comments on the early draft of the manuscript.